\documentclass[conference]{IEEEtran}
\IEEEoverridecommandlockouts
\usepackage{cite}
\usepackage{listings}
\usepackage{amsmath,amssymb,amsfonts}
\usepackage{algorithmic}
\usepackage{graphicx}
\usepackage{textcomp}
\usepackage{xcolor}
\usepackage{url}
\usepackage{balance}
\usepackage{hyperref}
\usepackage{tcolorbox}

\definecolor{dkgreen}{HTML}{006400}
\definecolor{gray}{rgb}{0.5,0.5,0.5}
\definecolor{azure}{HTML}{142086}

\newcommand{\etal}{\textit{et al}. }

\newcommand{\eg}{\textit{e}.\textit{g}., }

\def\BibTeX{{\rm B\kern-.05em{\sc i\kern-.025em b}\kern-.08em%
    T\kern-.1667em\lower.7ex\hbox{E}\kern-.125emX}}

\begin{document}

\title{AI-Powered Commit Explorer (APCE)}


\author{
    \IEEEauthorblockN{
        Yousab Grees\IEEEauthorrefmark{1},
        Polina Iaremchuk\IEEEauthorrefmark{1},
        Ramtin Ehsani\IEEEauthorrefmark{2},
        Esteban Parra\IEEEauthorrefmark{1},
        Preetha Chatterjee\IEEEauthorrefmark{2},
        Sonia Haiduc\IEEEauthorrefmark{3}
    }
    \IEEEauthorblockA{\IEEEauthorrefmark{1}Department of Mathematics, Computer Science, and Data Science, Belmont University, Nashville, TN, USA\\
    \{yousab.grees@bruins.belmont.edu, polina.iaremchuk@bruins.belmont.edu, esteban.parrarodriguez@belmont.edu\}}
    
    \IEEEauthorblockA{\IEEEauthorrefmark{2}Department of Computer Science, Drexel University, Philadelphia, PA, USA \\
    \{ramtin.ehsani@drexel.edu, preetha.chatterjee@drexel.edu\}}
    
    \IEEEauthorblockA{\IEEEauthorrefmark{3}Department of Computer Science, Florida State University, Tallahassee, FL, USA\\
    shaiduc@fsu.edu}
}



\maketitle

\begin{abstract}
Commit messages in a version control system provide valuable information for developers regarding code changes in software systems. 
Commit messages can be the only source of information left for future developers describing what was changed and why. However, writing high-quality commit messages is often neglected in practice. Large Language Model (LLM) generated commit messages have emerged as a way to mitigate this issue. We introduce the AI-Powered Commit Explorer (APCE), a tool to support developers and researchers in the use and study of LLM-generated commit messages. APCE gives researchers the option to store different prompts for LLMs and provides an additional evaluation prompt that can further enhance the commit message provided by LLMs. APCE also provides researchers with a straightforward mechanism for automated and human evaluation of LLM-generated messages. Demo link \url{https://youtu.be/zYrJ9s6sZvo}

\end{abstract}


\begin{IEEEkeywords}
Large Language Models, GitHub, Automated Commit Messages, AI4SE, Code Summarization
\end{IEEEkeywords}

\section{Introduction}
\label{sec:introduction}
As part of the routine software maintenance and evolution process of large software systems, developers often need to reference previous changes to fix bugs or implement new features. Changes to software artifacts are tracked through Version Control Systems (e.g., Git). Git allows developers to describe the changes to the software artifacts via a commit message containing a textual description of the changes within the commit as well as the rationale behind those changes~\cite{Zhang2024b}. 

Commit messages play a critical role in developers' understanding and communicating code changes and software maintenance. In particular, commit messages can be the only source of information left for future developers in long-lived projects~\cite{Li2023}. Therefore, high-quality commit messages are crucial for the long-term maintenance and evolution of any software project. However, writing high-quality commit messages is often neglected in practice, leading to commit messages that are incomplete, ambiguous, non-informative, difficult to understand, or empty~\cite{Tian2022, Dyer2013, Li2023}.

Automatic commit message generation aims to leverage computational methods to develop approaches to support developers in their software maintenance process by automatically generating meaningful commit messages that provide them information regarding \textit{what?} and \textit{why?} a set of code changes where made~\cite{Cortes2014, Li2023}. Automated commit generation approaches can be classified into four main categories: rule-based, retrieval-based, learning-based, and hybrid~\cite{Dong2022}. 

The rise of Large Language Models (LLMs) and multi-agent frameworks composed of multiple LLM-based agents that interact and collaborate to solve complex problems or achieve goals beyond the capability of any single agent, has led to a growing body of research on using LLMs in various software engineering tasks~\cite{White2023, Mohamed2024, Sallou2024, Shehata2024, He2025}. Among them are LLM-based approaches for automatically generating commit messages~\cite{Tao2022, Zhang2024b, Zhang2024a, Xue2024}. As these newer approaches continue to evolve and demonstrate promising performance~\cite{Zhang2024b}, it is necessary to evaluate the newer approaches quickly and seamlessly, ensuring timely validation and comparison.

In this paper, we introduce the AI-Powered Commit Explorer (APCE), a tool aimed at assisting researchers in the generation and evaluation of commit messages using LLM-based approaches. APCE serves two primary purposes. First, APCE provides seamless integration into GitHub repositories, enabling the automatic generation of high-quality commit messages that describe both the what and the why of code changes. By selecting any commit within a GitHub-hosted repository, developers can use APCE to obtain a synthesized, contextual message generated by one or multiple LLM-based approaches, aiding them in their code understanding and long-term maintainability of the system. Second, APCE facilitates and streamlines empirical analysis and evaluation of commit messages generated by new LLM-based approaches with a built-in evaluation module. The evaluation module compares automatically generated commit messages to human-written counterparts and computes standard evaluation metrics (\eg BLEU, ROUGE-L, and METEOR) ~\cite{Roy2021, Tao2022}. Furthermore, the evaluation module supports the empirical evaluation of new automated commit message generation techniques by collecting users' feedback on the quality of the automated commit messages (\eg completeness, consistency, and informativeness). The full implementation of APCE is available on GitHub\footnote{\url{https://github.com/yousabg/AI-Powered-Commit-Explorer}} and our replication package\footnote{\url{https://figshare.com/s/0f1f15af0ecb5aeedee2}}.

The rest of this paper is structured as follows. Section \ref{sec:relatedwork} will explain the works that inspired this tool. Section \ref{sec:architecture} presents the architecture and design of our tool, and Section \ref{sec:availability} outlines the tool's availability. Finally, section \ref{sec:futurework} will conclude the paper with future work and limitations.

\section{Related Work}
\label{sec:relatedwork}


In this section, we briefly discuss existing work in automated commit message generation. Approaches for automatic generation of commit messages can be broadly classified into generation-based and retrieval-based methods\cite{Cortes2014, Liu2019, Xu2019, Liu2020}. Retrieval-based approaches use Natural Language Processing (NLP) and Information Retrieval (IR) techniques to pick the most similar words or messages from a large dataset, whereas generation-based approaches use machine learning and large datasets to predict the words that the commit message should contain \cite{Zhang2024b}.

One of the first retrieval-based tools proposed to address this problem was ChangeScribe\cite{Cortes2014}. ChangeScribe utilized NLP to compare the code differences of a commit. Then, it formulates a short sentence based on that difference. Since then, several methods have been built to improve upon this work. For example, Liu et al. \cite{Liu2018} proposed NNGen, which uses the nearest neighbors algorithm to retrieve the most similar code diff from a training set and reuse its commit message. Additionally, Wang et al. \cite{Wang2021} introduced CoRec that combines retrieval and machine learning approaches. CoRec uses a decay sampling strategy during training to shift from ground truth to model-generated inputs and leverages the most similar
commit to refine its output.

More recently, generation-based methods using neural architectures and pre-trained models have emerged~\cite{Dong2022, Liu2023, Zhang2024a}. FIRA, an approach that utilizes fine-grained graph representations of code changes and a graph-neural network encoder with a transformer-based decoder improved by a dual copy mechanism allowing flexible access to sub-token and integral token information\cite{Dong2022}. The CCRep, a newer technique introduced by Liu \etal \cite{Liu2023}, leverages a pre-trained model to encode code changes and generate a commit message. Experiments using models such as $CCRep^{token}$, $CCRep^{line}$, and $CCRep^{hybrid}$ demonstrated improvement in the quality of generated messages. Zhang \etal \cite{Zhang2024a} conducted a study evaluating commit message generation using popular open-source and closed-source LLMs such as ChatGPT and Llama. The two-phase evaluation consisted of a zero-shot setup, where prompts from a dataset of commit messages were sent through ChatGPT and Llama2, and then evaluated by BLEU and Rouge-L. Then, the messages were shuffled and evaluated by the two researchers. Surprisingly, LLM-generated messages were preferred over human-written ones in most cases—human messages were favored only 13.1\% of the time.

APCE contributes to the study of automated commit generation by providing an easy-to-set-up framework to streamline the evaluation of LLM-based commit generation approaches.

\section{Architecture}
\label{sec:architecture}

APCE is a web-based tool that supports the generation and evaluation of commit messages. APCE uses a web application architecture in which the front-end client is built using \textit{Next.js}\footnote{\url{https://nextjs.org/}}, a JavaScript web development framework, the back-end is hosted on a Flask\footnote{\url{https://flask.palletsprojects.com/en/stable/}} server, a lightweight Python web framework for building APIs and handling HTTP requests, and using a MySQL\footnote{\url{https://www.mysql.com/}} database to store the data.

\begin{figure}[ht!]
    \centering
    \includegraphics[width=\linewidth]{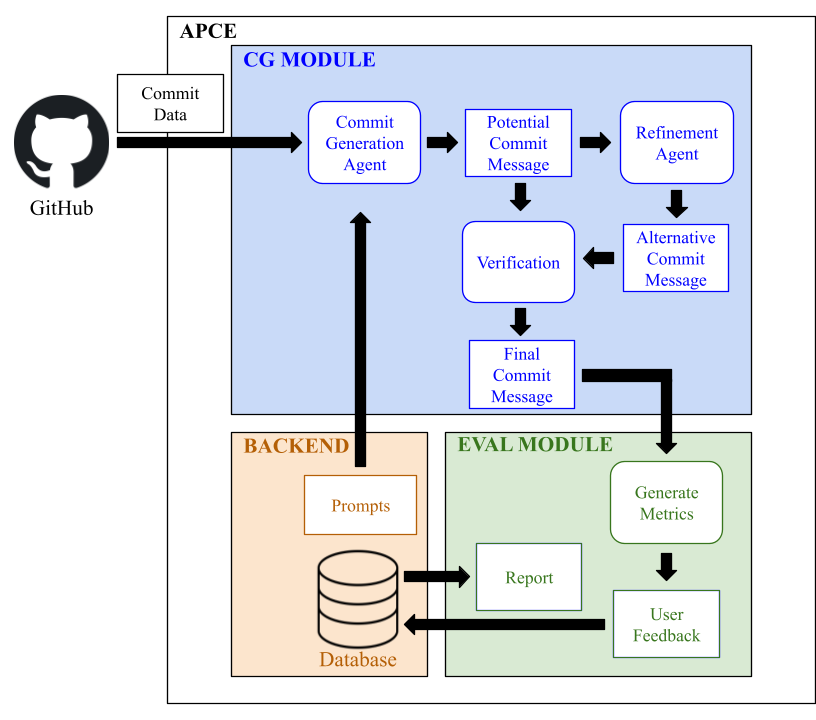}
    \caption{APCE Architecture}
    \label{fig:architecture}
\end{figure}

APCE has two modules: (i) a commit generation module and (ii) an evaluation module, as shown in Figure~\ref{fig:architecture}.  

The commit generation module (CG MODULE) uses a Multi-agent framework \cite{He2025} to generate a commit message. In particular, APCE leverages two agents (i.e., a commit Generation Agent and a Refinement Agent) that engage in a multi-turn self-collaboration process. 

The evaluation module (EVAL MODULE) supports the following functionality: approach management, consent form, data collection, user interaction, evaluation metric computation, and reporting, which allows researchers to work purely on analyzing results rather than building an infrastructure.

The APCE provides a consent form when a participant first loads the tool. Currently, the sample consent form is tailored to Belmont University, since the author of the tool conducted research at that university. The consent form used can be replaced to meet the researcher's specific institutional needs. The GitHub credentials require a user to make a choice by clicking presented 'Accept' button to proceed with a tool.

\begin{figure*}[ht!]
    \centering
    \includegraphics[width=\linewidth]{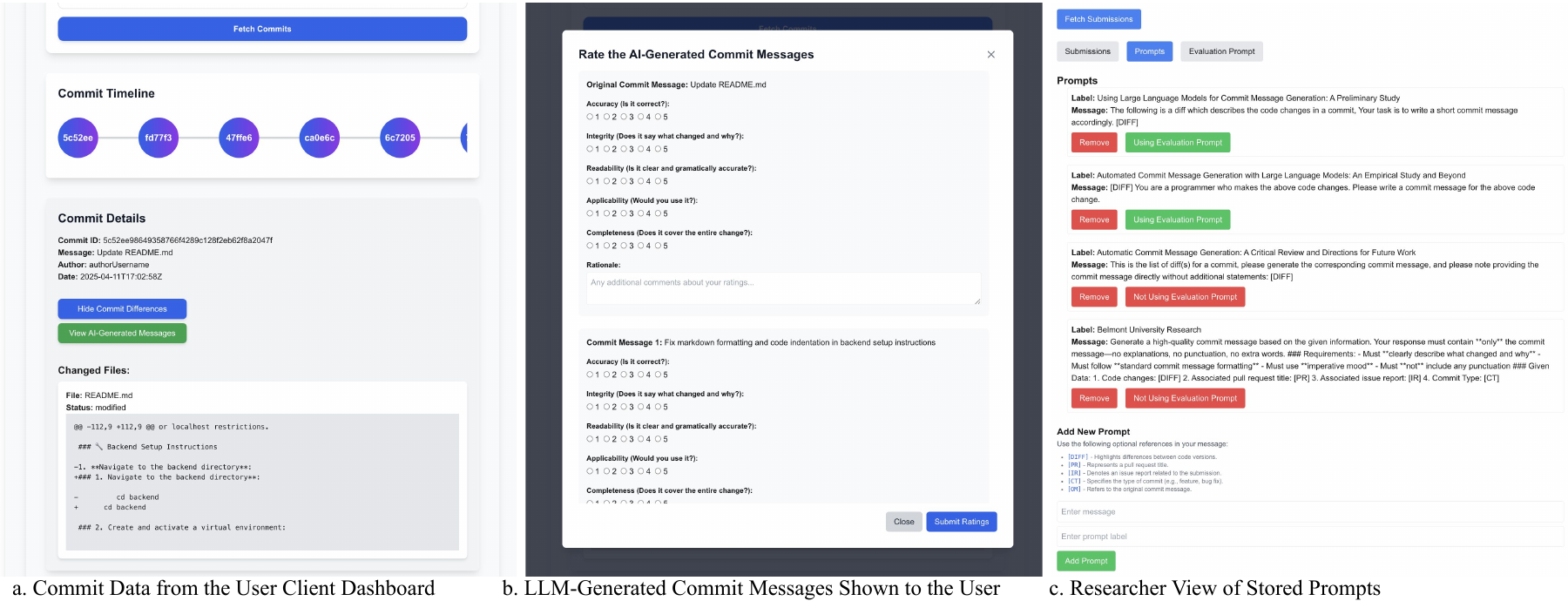}
    \caption{APCE GUI}
    \label{fig:GUI}
    \vspace{-0.2cm}
\end{figure*}

Assuming consent is accepted, the user is then prompted to share their GitHub token and username. The GitHub token is a digital authentication key that is used to access the OctoKit API\footnote{\url{https://github.com/octokit}}, a collection of client libraries, that simplifies interaction with the GitHub API, and pulls the user's repositories and commits. For the sake of secure and correct final commit message generation, as emphasized in the consent form, personal data that can link back to the user is not collected, besides the data specific to the actual commit. 

\subsection{Commit Generation Module}
The commit generation module allows the user to select a repository. After selecting a repository, the user is presented with a list of all commits associated with that repository in a node timeline as shown in Figure \ref{fig:GUI}a. When selecting a commit node, APCE populates with the details about the commit and enables a button to 'View AI Generate Messages'. When the user clicks on the button, the commit details will be replaced with a loading log while APCE generates a commit message for the selected commit using each of the configured commit generation approaches.

APCE uses two agents in the commit generation process. The commit generation agent uses a prompt to generate a commit message, while the refinement agent assesses the commit message generated by the commit generation agent. The refinement agent can be enabled or disabled for any approach. The default prompt used by the refinement agent is shown in Figure \ref{fig:RefinementPrompt}. The two-agent framework allows APCE to be easily modified to support any existing and future LLM-based commit generation approaches by adding or removing a prompt to the backend (See Section \ref{sec:ApproachesAPCE} for details on how to set up approaches in APCE).

\begin{figure}[h]
\begin{tcolorbox}[]
Evaluate the commit message below.

If it fully meets all criteria, reply only with the exact same commit message. 

If it does not fully meet all criteria, reply only with a corrected commit message.  

Your response must contain nothing else—no explanations, no punctuation, no extra words.

Criteria:
\begin{itemize}
  \item Must be less than 72 characters
  \item Must use imperative mood (e.g., ``Fix bug" instead of ``Fixed bug")
  \item Must clearly describe the change
  \item Must not include explanations or reasoning
  \item May describe multiple changes
\end{itemize}

Commit message to evaluate: ``[MESSAGE]"
\end{tcolorbox}
\vspace{-0.2cm}
    \caption{Refinement Agent prompt}
    \label{fig:RefinementPrompt}
\end{figure}

For each of the approaches configured in APCE: 1) the commit generation agent will first generate a potential commit message using the corresponding prompt, 2) the refinement agent will asses the potential commit message generated by the commit generation agent and generate an alternative commit message if the potential commit message does not follow the format of a commit message, 3) APCE will compare the potential and the alternative commit messages and select one of the two.

If an alternative prompt is returned by the refinement agent, APCE will compare both the potential and the alternative commit messages and select one of the two. First, if either response is not a valid commit message based on the criteria, then the other response is chosen. If both are not valid, an error is generated for this message. If both are valid, then it checks if any of the generated commit messages are greater than 72 characters. If one of them is, then the client will choose the other commit message, since we prefer a commit message that is less than 72 characters for readability between tools and terminals\cite{beams2014git}. Lastly, if both commit messages are less than 72 characters, then the verification will choose the longer commit message. If the user configures an approach to skip the refinement agent, APCE would only check the validity of the initial commit message by ensuring that the response is not an error and that the response is less than 200 characters. APCE uses the 200-character limit to prioritize returning a commit message over an error and avoiding overly verbose messages.

APCE uses the unified API service OpenRouter\footnote{\url{https://openrouter.ai/docs/quickstart}} to support access to various LLM models. In the default implementation, APCE utilizes the DeepSeek\footnote{\url{https://chat.deepseek.com}} model on OpenRouter, which has consistently delivered the most accurate responses among free-tier models. To use OpenRouter, the developer or researcher must set up a free account and generate an API key. This key should be added to the code-base via a .env.local file in the following format:

\begin{lstlisting}[language=bash]
NEXT_PUBLIC_OPENROUTER_API_KEY=<your_api_key_here>
\end{lstlisting}

If any errors occurred during an API call to the LLM to generate a commit message for a given approach, APCE will re-attempt after 5000ms. After 3 unsuccessful tries, APCE will return an error for that message.

            
            
            

Although the default LLM in APCE is deepSeek-R1 via OpenRouter, the tool can be customized to fit the needs of other researchers by changing the LLM model and provider by modifying the \textit{api.js} file. In particular, if a researcher wants to change the LLM model used, then only the body of the API needs to be changed in the \textit{getDeepSeekResponse()} method. On the other hand, if a researcher wants to add a different API service, they will have to write a new method to replace the existing \textit{getDeepSeekResponse} method (for example, using a cron job or a simple js compiler).

\subsection{Configuring Commit Generation Approaches in APCE}
\label{sec:ApproachesAPCE}
Using the research view (see Figure \ref{fig:GUI}c), the user can configure one or more LLM-based commit generation approaches to be used by the commit generation module. The following optional references can be added to the approach to include the corresponding information from the commit itself:

        \begin{itemize}
            \item \texttt{[DIFF]} – Differences between code versions.
            \item \texttt{[PR]} – Title of the pull request.
            \item \texttt{[IR]} – The issue report related to the submission.
            \item \texttt{[CT]} – The type of commit (e.g., feature, bug fix).
            \item \texttt{[OM]} – The original commit message.
        \end{itemize}

The generation prompt used by the default commit generation approach configured in APCE can be seen in Figure \ref{fig:Prompt}. Furthermore, as seen in Figure \ref{fig:GUI}c, APCE is already configured to use the approaches by Xue et al. \cite{Xue2024} and Zhang et al. \cite{Zhang2024a}.

\subsection{Evaluation Module}
The evaluation module is created to support researchers in performing studies to evaluate new automated commit generation approaches. 

APCE will evaluate the similarity between the original commit message and the LLM-generated message(s) by computing the BLEU, METEOR, and ROUGE-L~\cite{Xue2024} evaluation metrics for each of the LLM-generated messages, well-known summarization metrics used to assess the quality of computer-generated commit messages\cite{Zhang2024a}. However, there are cases where the original message may hold truth, but  have minor overlap with the LLM-generated message, even if the meaning remains the same \cite{Zhang2024b}. Therefore, APCE supports the collection of human evaluation feedback.

\begin{figure}[h]
\begin{tcolorbox}[]
\textit{\textbf{Prompt:}} \textit{Generate a high-quality commit message based on the given information. Your response must contain only the commit message—no explanations, no punctuation, no extra words.}
                
\textit{\#\#\# Requirements:
                - Must clearly describe what changed and why  
                - Must follow standard commit message formatting  
                - Must use imperative mood  
                - Must not include any punctuation}
                
                \#\#\# Given Data: \\
                1. Code changes: [DIFF]  \\
                2. Associated pull request title: [PR] \\ 
                3. Associated issue report: [IR]  \\
                4. Commit Type: [CT]

                \textbf{Approach Name:} {\textit{Default}} \\
                \textbf{Refinement:} \texttt{true}
\end{tcolorbox}
    \vspace{-0.2cm}
    \caption{Commit Generation Agent Prompt Example}
    \label{fig:Prompt}
\end{figure}

When the user clicks the "View AI-Generated Messages" button, a modal pop-up window appears for the user to rate and provide feedback for each of the commit messages generated by the different approaches. The modal shows the user the original message, followed by AI-generated commit messages displayed in a shuffled order with arbitrary indexes, as shown in Figure~\ref{fig:GUI}b. In particular, the user is asked to rate each LLM-generated message on a 5-point Likert scale using five quality criteria derived from existing work \cite{Cortes2014, Xue2024}. The criteria are: \textit{accuracy} (Is the commit message correct?), \textit{integrity} (Does it explain what changed and why?), \textit{readability} (Is the commit message clear and free of grammatical errors?), \textit{applicability} (Would other developers use the same commit message?), and \textit{completeness} (Does the commit message cover all the changes?). 

Lastly, for each LLM-generated message, the user is asked to include a rationale for their ratings. Once a user successfully submits the ratings, the submission is stored in the database. 

For each submission, APCE stores the following attributes:
\begin{itemize}
        \item A unique submission ID
        \item The issue report of the GitHub commit
        \item The Commit ID
        \item The Commit Type
        \item The original message associated with the GitHub commit
        \item The title of the pull request of the GitHub commit
        \item The timestamp of the GitHub commit
        \item The files associated with the GitHub commit:
        \begin{itemize}
            \item Filename
            \item File status (e.g., added, modified, deleted)
            \item Number of additions
            \item Number of changes
            \item Number of deletions
        \end{itemize}
        \item The ratings associated with the submission:
        \begin{itemize}
            \item The generated commit message
            \item The Name of the approach used
            \item Whether the prompt generated a successful commit message
            \item Whether the refinement prompt was used
            \item User's ratings
            \item Rationale for the ratings
            \item Evaluation metric scores (BLEU, METEOR, ROUGE-L)
        \end{itemize}
\end{itemize}

During the evaluation, the researcher can access stored submissions using the password-protected research view, shown in Figure~\ref{fig:GUI}c. The research view shows the submissions, prompts, approaches, and refinement prompt. In the \itshape{Prompts} \normalfont section, the user can add or remove approaches, or adjust the refinement prompt settings of the chosen approach.


\section{Availability}
\label{sec:availability}
More about APCE can be found in the tool's GitHub repository, which contains (i) setup and configuration instructions, (ii) the source code, and (iii) the architecture description. 

\section{Limitations and Future Work}
\label{sec:futurework}


One limitation of APCE is that a large number of API calls can significantly slow down the tool. As noted earlier, performance drops as the commit diff gets larger, which worsens the commit message quality. Furthermore, its usability can be hindered for projects with longer commit histories due to GitHub’s hourly rate limit for API usage. In the future, we will work on implementing asynchronous processing, caching, and queuing to help reduce the delay and increase performance when multiple users are using the system concurrently.

Future work includes implementing a feature to import commit message datasets, allowing researchers to perform a bulk analysis of data to will speed up the non-human evaluation of approaches and incorporating additional evaluation metrics and reports.

Currently, researchers are the primary intended users and beneficiaries of APCE. However, developers can use the commit generation module and skip the evaluation. Future iterations of APCE aim to separate the commit generation and evaluation modules, making the tool more accessible and useful for developers.


\section*{Acknowledgment}
Esteban Parra and Polina Iaremchuk were supported in part by Belmont University through the Summer Undergraduate Research Fellowships in the Sciences (SURFS) program.

\balance

\bibliographystyle{IEEEtranS}
\bibliography{references}
\end{document}